\documentclass[english,prl,aps,twocolumn,superscriptaddress]{revtex4-1}
\usepackage[T1]{fontenc}

\usepackage{array}
\usepackage{booktabs}
\usepackage{multirow}
\usepackage{babel}

\usepackage{bm}
\usepackage{comment}
\usepackage{amssymb}
\usepackage{amsfonts}
\usepackage{graphicx}
\usepackage{hyperref} \hypersetup{colorlinks=true,citecolor=red,linkcolor=red,urlcolor=red} % HYPERLINKS
\usepackage{color}
\usepackage{longtable}

\usepackage[dvipsnames]{xcolor}

\begin{document}

%\title{An Automated Application Framework to Model Disordered Materials Based on a High Throughput First Principles Approach}
\title{Modeling Disordered Materials with a High Throughput {\it ab-initio} Approach}

\author{Kesong Yang}
\affiliation{\small Department of NanoEngineering, University of California San Diego, La Jolla, California 92093-0448, USA}
\author{Corey Oses}
\affiliation{\small Center for Materials Genomics, Duke University, Durham, North Carolina 27708, USA}
\author{Stefano Curtarolo}
\affiliation{\small Materials Science, Electrical Engineering, Physics and Chemistry, Duke University, Durham, North Carolina 27708, USA}
\affiliation{email: stefano@duke.edu}

\begin{abstract}
Predicting material properties of disordered systems remains a long-standing and formidable challenge in rational materials design.
To address this issue, we introduce an automated software framework capable of modeling partial occupation within 
disordered materials using a high-throughput (HT) first principles approach.
At the heart of the approach is the construction of supercells containing a virtually equivalent stoichiometry to the disordered material.
All unique supercell permutations are enumerated and material properties of each are determined via HT electronic structure calculations.
In accordance with a canonical ensemble of supercell states, the framework evaluates ensemble average properties of the system as a function
of temperature.
As proof of concept, we examine the framework's final calculated properties of 
a zinc chalcogenide (ZnS$_{1-x}$Se$_x$), a wide-gap oxide semiconductor (Mg$_{x}$Zn$_{1-x}$O), and an iron alloy (Fe$_{1-x}$Cu$_{x}$)
at various stoichiometries.

\end{abstract}
\maketitle

\section{Introduction}
Crystals are characterized by their regular, repeating structures.
Such a description allows us to reduce our focus from the macroscopic material to a microscopic subset of
unique atoms and positions.
A full depiction of material properties, including mechanical, electronic, and magnetic features, 
follows from an analysis of this primitive lattice.
First principles quantum mechanical calculations have been largely successful in reproducing ground state properties of 
perfectly ordered crystals~\cite{DFT,Hohenberg_PR_1964}.
However, such perfection does not exist in nature.
Instead, crystals display a degree of randomness, or disorder, in their lattices.
There are several types of disorder; including topological, spin, substitutional, and vibrational~\cite{Elliott_PoAM_1990}.
This work focuses on substitutional disorder, in which crystallographically equivalent sites of a crystal are not uniquely or fully occupied.
Rather, each site is characterized by a statistical, or partial, occupation.
Such disorder is intrinsic in many technologically significant systems, including those used in fuel cells~\cite{Xie_ACB_2015}, solar cells~\cite{Kurian_JPCC_2013}, 
high-temperature superconductors~\cite{Bednorz_ZPBCM_1986,Maeno_Nature_1994}, low thermal conductivity 
thermoelectrics~\cite{Winter_JACerS_2007}, imaging and communications devices~\cite{Patra_JAP_2012}, as well as promising 
rare-earth free materials for use in free sensors, actuators, energy-harvesters, and spintronic devices~\cite{Wang_SR_2013}.
Hence, a comprehensive computational study of substitutionally disordered materials at the atomic scale is of paramount importance for 
optimizing key physical properties of materials in technological applications.

Unfortunately, structural parameters with partial occupancy cannot be used directly in first principles calculations---a 
significant hindrance for computational studies of disordered systems.
Therefore, additional efforts must be made to model this disorder.
One method relies on the reformulation of the disordered system into an average effective virtual compound, 
\textit{i.e.}, virtual crystal approximation (VCA)~\cite{Nord_1931_AP_VCA, Vanderbilt_2000_PRB_VCA}.
In the VCA approach, each disordered site is treated as a virtual atom possessing attributes that are the 
compositional-weighted average of the actual occupants.
An advantage of the VCA approach is that the computational cost of a disordered material is comparable to that of an ordered material.
However, this approach neglects local distortional effects around the partially occupied sites---obscuring fine features of the overall structure.

An alternative method is the mean-field-type coherent potential approximation (CPA)~\cite{Soven_PhysRev_1967}, often implemented 
within the Korringa-Kohn-Rostoker (KKR) \cite{Korringa1947392, Kohn_1954_PhysRev, Stocks_PRL_1978} multiple scattering formalism for 
enhanced model accuracy~\cite{Faulkner_PM_2006}.
This approach simulates interactions between electrons and the potential via propagators (Green's functions) and approximates the 
varying potentials of random alloys by introducing an effective medium potential from a perfectly ordered lattice~\cite{Faulkner_PMS_1982}.
Like any mean-field approximation, the CPA description is a single-site approximation and thus unable to resolve short range order effects,
such as fine resonance structure in the electronic density of states (DOS).

These approaches, among others~\cite{Zunger_PRL_1990,Shan_PRL_1999,Popescu_PRL_2010},
grapple with the legitimacy of attributing electronic band structure properties to random alloys, which 
exhibit no translational long range order.
This work takes a different approach, reformulating the issue into one of a statistical nature.
While not all material properties may be garnered through this reformulation, those that are accessible are of significant practical importance,
including the DOS, band gap energy E$_{gap}$, and  magnetic moment M.

A rigorous statistical treatment of substitutional disorder at the atomic scale requires utility of large ordered supercells 
containing a composition consistent with the compound's stoichiometry~\cite{Habgood_PCCP_2011,Haverkort_ArXiv_2011}.
However, the computational cost of such large supercell calculations has traditionally inhibited their use.
Fortunately, the emergence of high-throughput (HT) computational techniques \cite{nmatHT} coupled with the exponential growth of computational power is 
now allowing the study of disordered systems from first principles~\cite{MGI_OSTP}.

Herein, we present an approach to perform such a treatment working within the HT computational framework AFLOW~\cite{AFLOW_STANDARD,aflowPAPER}.
We highlight three novel and attractive features central to this method:  complete implementation into an automatic high throughput framework (optimizing speed without
mitigating accuracy), utility of a novel occupancy optimization algorithm, and use of the Universal Force Field method \cite{Rappe_1992_JCAS_UFF} 
to reduce the number of DFT calculations needed per system.

\section{Methodology}

This section details the technicalities of representing a partially occupied disordered system as a series of unique supercells.
Here is an outline of the approach:

(1)	For a given disordered material, optimize its partial occupancy values and determine the size of the derivative superlattices.

(2) 
(\textit{a}) Use the superlattice size \textit{n} to generate a set of unique derivative superlattices and corresponding sets of 
unique supercells with the required stoichiometry.
(\textit{b}) Import these non-equivalent supercells into the automatic computational framework AFLOW for HT 
first principles electronic structure calculations.

(3) Obtain and use the relative enthalpy of formation to calculate the equilibrium probability of each 
supercell as a function of temperature \textit{T} according to the Boltzmann distribution.

(4) Determine the disordered system's material properties through ensemble averages of the properties calculated for each supercell.
Specifically, we will be calculating the system's density of states (DOS), band gap energy E$_{gap}$, and magnetic moment M.

In the following sections, we will refer to a model disordered system, Ag$_{8.733}$Cd$_{3.8}$Zr$_{3.267}$, to illustrate the technical procedures
mentioned above.
This disordered system has two partially occupied sites:  one shared between silver and zirconium, and another shared between
cadmium and a vacancy.
Working within the AFLOW framework~\cite{Setyawan_2010_CMS.299}, we have designed a simple structure file for partially occupied systems.
Adapted from VASP's POSCAR~\cite{VASP_PRB}, the PARTCAR contains within it a description of lattice parameters and a 
list of site coordinates and occupants, along 
with a concentration tolerance (explained in the next section), and (partial) occupancy values for each site.
To see more details about this structure or its PARTCAR, please see the Supplementary Materials.

\subsection{Determine superlattice size}
In order to fully account for the partial occupancy of the disordered system, we would need to generate a set of superlattices of 
a size corresponding to the lowest common denominator of the fractional partial occupancy values.
With partial occupancy values of 0.733 (733/1000) and 0.267 (267/1000) in the disordered system Ag$_{8.733}$Cd$_{3.8}$Zr$_{3.267}$, 
we would need to construct superlattices of size 1000.
Not only would we be working with correspondingly large supercells (16,000 atoms per supercell in our example), 
but the number of unique supercells in the set would be substantial.
This would extend well beyond the capability of first principles calculations, and thus, is not practical.
It is therefore necessary to optimize the partial occupancy values to produce an appropriate superlattice size.

\begin{table}
\caption{
Evolution of the algorithm used to optimize the partial occupancy values and superlattice size for the disordered system
Ag$_{8.733}$Cd$_{3.8}$Zr$_{3.267}$. 
\textit{f$_i$} indicates the iteration's choice fraction for each partially occupied site, (\textit{i} = 1, 2, 3, \ldots);
\textit{e$_i$} indicates the error between the iteration's choice fraction and the actual partial occupancy value.
\textit{$e_{max}$} is the maximum error of the system.
}
\vspace{2mm}

{\footnotesize
\begin{tabular*}{0.95\columnwidth}{@{\extracolsep{\fill}}ccccccccc}
\toprule 
\multirow{2}{*}{\textit{n}\'} & \multicolumn{2}{c}{Occup. 1 (Ag)} & \multicolumn{2}{c}{Occup. 2 (Zr)} & \multicolumn{2}{c}{Occup. 3 (Cd)} & \multirow{2}{*}{\textit{$e_{max}$} } & \multirow{2}{*}{\textit{n}}\tabularnewline
\cmidrule{2-7} 
 & \textit{f$_{1}$}  & \textit{e$_{1}$}  & \textit{f$_{2}$}  & \textit{e$_{2}$}  & \textit{f$_{3}$}  & \textit{e$_{3}$}  &  & \tabularnewline
\midrule 
1	& 1/1   & 0.267  & 0/1  & 0.267  & 1/1   & 0.2    & 0.267  & 1\tabularnewline
\midrule                                              
2	& 1/2   & 0.233  & 1/2  & 0.233  & 2/2   & 0.2    & 0.233  & 2\tabularnewline
\midrule                                              
3	& 2/3   & 0.067  & 1/3  & 0.067  & 2/3   & 0.133  & 0.133  & 3\tabularnewline
\midrule                                              
4	& 3/4   & 0.017  & 1/4  & 0.017  & 3/4   & 0.05   & 0.05   & 4\tabularnewline
\midrule                                              
5	& 4/5   & 0.067  & 1/5  & 0.067  & 4/5   & 0      & 0.067  & 5\tabularnewline
\midrule                                              
6	& 4/6   & 0.067  & 2/6  & 0.067  & 5/6   & 0.033  & 0.067  & 6\tabularnewline
\midrule                                              
7	& 5/7   & 0.019  & 2/7  & 0.019  & 6/7   & 0.057  & 0.057  & 7\tabularnewline
\midrule                                              
8	& 6/8   & 0.017  & 2/8  & 0.017  & 6/8   & 0.05   & 0.05   & 4\tabularnewline
\midrule                                              
9	& 7/9   & 0.044  & 2/9  & 0.044  & 7/9   & 0.022  & 0.044  & 9\tabularnewline
\midrule                                              
10	& 7/10  & 0.033  & 3/10 & 0.033  & 8/10  & 0      & 0.033  & 10\tabularnewline
\midrule                                              
11	& 8/11  & 0.006  & 3/11 & 0.006  & 9/11  & 0.018  & 0.018  & 11\tabularnewline
\midrule                                              
12	& 9/12  & 0.017  & 3/12 & 0.017  & 10/12 & 0.033  & 0.033  & 12\tabularnewline
\midrule                                              
13	& 10/13 & 0.036  & 3/13 & 0.036  & 10/13 & 0.031  & 0.036  & 13\tabularnewline
\midrule                                              
14	& 10/14 & 0.019  & 4/14 & 0.019  & 11/14 & 0.014  & 0.019  & 14\tabularnewline
\midrule                                              
15	& 11/15 & 0.00003 & 4/15 & 0.00003 & 12/15 & 0      & 0.00003 & 15\tabularnewline
\bottomrule
\end{tabular*}
}
\end{table}

We demonstrate utility of an efficient algorithm to calculate the optimized partial occupancy values and corresponding superlattice size 
with our example disordered system Ag$_{8.733}$Cd$_{3.8}$Zr$_{3.267}$ in Table I.
For convenience, we refer to the algorithm's iteration step as \textit{n}\'\ , 
the superlattice index, and \textit{n} as the superlattice size.
Quite simply, the algorithm iterates, increasing the superlattice index from 1 to \textit{n}\'\, until the optimized partial occupancy values reach the required accuracy.
At each iteration, we generate a fraction for each partially occupied site, all of which have the common denominator \textit{n}\'\ .
The numerator is determined to be the integer that reduces the overall fraction's error relative to the actual site's fractional partial occupancy value.
The superlattice size corresponds to the lowest common denominator of the irreducible fractions (\textit{e.g.} see iteration step 8).
The maximum error among all of the sites is chosen to be the accuracy metric for the system.

For the disordered system Ag$_{8.733}$Cd$_{3.8}$Zr$_{3.267}$, given a tolerance of 0.01, the calculated superlattice size is 15
(240 atoms per supercell).
By choosing a superlattice with a nearly equivalent stoichiometry as the disordered system, we have reduced
our supercell size by over a factor of 60 and entered the realm of feasibility with this calculation.

Quite expectedly, we see that the errors in partial occupancy values calculated for 
silver and zirconium are the same, as they share the same site.
The same holds true for cadmium and its vacant counterpart (not shown).
Therefore, the algorithm only needs to determine one choice fraction per site, instead of per occupant (as shown).
Such an approach reduces computational costs by guaranteeing that only the smallest supercells (both in number and size) 
with the lowest tolerable error in composition are funneled into our HT first principles calculation framework.

\subsection{Unique supercells generation}
%\textit{Production of all non-equivalent supercells:} 
With the optimal superlattice size \textit{n}, the unique derivative superlattices of the disordered system can be generated using 
Hermite Normal Form (HNF) matrices~\cite{Gus_2008_PRB_ENUM}.
Each HNF matrix generates a superlattice of a size corresponding to its determinant, $n$.
There exists many HNF matrices with the same determinant, each creating a variant superlattice.
For each unique superlattice, we generate a complete set of possible supercells with the required stoichiometry by exploring all
possible occupations of partially occupied sites.
However, not all of these combinations are unique---nominally warranting an involved structure comparison analysis that becomes
extremely time consuming for large supercells~\cite{Gus_2008_PRB_ENUM}.
Instead, we identify duplicates by estimating the total energy of each supercell in a HT manner based on the Universal Force Field (UFF)
method~\cite{Rappe_1992_JCAS_UFF}.
This classical molecular mechanics force field approximates the energy of a structure by considering its composition, 
connectivity, and geometry, for which parameters have been tabulated.
Only supercells with the same total energy are structurally compared and potentially treated as duplicate structures to be discarded, if necessary.
The count of duplicate structures determines the degeneracy of the structure.
Only non-equivalent supercells are imported into the automatic computational framework AFLOW for HT 
quantum mechanics.

\subsection{Supercell equilibrium probability calculation}
The unique supercells representing a partially occupied disordered material are labeled as {$S_1$, $S_2$, $S_3$, \ldots, 
$S_n$}. 
Their formation enthalpies (per atom) are labeled as {$H_{F,1}$, $H_{F,2}$, $H_{F,3}$, \ldots, $H_{F,4}$}, respectively.
The formation enthalpy of each supercell is automatically calculated from HT first principles calculations using the AFLOW 
framework~\cite{AFLOW_STANDARD,aflowPAPER}.
We take the supercell with the lowest formation enthalpy as a reference (ground state structure), and denote its formation enthalpy as $H_{F,0}$. 
The relative formation enthalpy of the \emph{i}th supercell is calculated as $\Delta {H_{F,i}} = {H_{F,i}} - {H_{F,0}}$
and characterizes its disorder relative to the ground state. 
The probability $P_i$ of the \emph{i}th supercell is determined by the Boltzmann factor:
\[{P_i} = \frac{{{g_ie^{ - \Delta {H_{F,i}}/{k_B}T}}}}{{\sum\limits_{i = 1}^n {{g_ie^{ - \Delta {H_{F,i}}/{k_B}T}}} }},\]
where $g_i$ is the degeneracy of the \emph{i}th supercell, 
$\Delta {H_{F,i}}$ is the relative formation enthalpy of the \emph{i}th supercell, 
$k_B$ is the Boltzmann constant, 
and \textit{T} is a virtual ``roughness'' temperature.
\textit{T} is not a true temperature \textit{per se}, but instead a parameter describing how much disorder has been 
statistically explored during synthesis.
To elaborate further, we consider two extremes in the ensemble average (ignoring structural degeneracy):  
1) $k_B T \lesssim \mathrm{max}\left(\Delta {H_{F,i}}\right)$
neglecting highly disordered structures $(\Delta {H_{F,i}} \ggg 0)$
as $T\to 0$, and 
2) $k_B T\ggg\mathrm{max}\left(\Delta {H_{F,i}}\right)$ 
representing the annealed limit ($T\to \infty$) in which all structures are considered equiprobable.
The probability $P_i$ describes the weight of the \emph{i}th supercell among the thermodynamically equivalent states of the disordered material
at equilibrium. 

\subsection{Ensemble average density of states, band gap energy, and magnetic moment}
With the calculated material properties of each supercell and its equilibrium probability in hand, the overall system properties
can be determined as ensemble averages of the properties calculated for each supercell.
This work focuses on the calculation of the ensemble average density of states (DOS), band gap energy E$_{gap}$, and magnetic moment M.
The DOS of the \emph{i}th supercell is labeled as $N_i$(E) and indicates the number of electronic states per energy interval.
The ensemble average DOS of the system is then determined by the following formula:

\[\mathrm{N}(E) = \sum\limits_{i = 1}^n {{P_i} \times {N_i}(E)}.\]

Additionally, a band gap $E_{gap,i}$ can be extracted from the DOS of each supercell.
In this fashion, an ensemble average band gap E$_{gap}$ can be calculated for the system.
It is important to note that standard density functional theory (DFT) calculations are limited to a description of the ground state~
\cite{DFT,Hohenberg_PR_1964}.
Subsequently, calculated excited state properties may contain substantial errors. 
In particular, DFT tends to underestimate the band gap~\cite{Perdew_IJQC_1985}.
Despite these known hindrances in the theory, we demonstrate in the next section that our framework is capable of predicting significant trends 
specific to the disordered systems.
As a bonus, the calculation of these results are performed in a high-throughput fashion.
It is expected that a more accurate, fine-grained description of the electronic structure in such systems will be obtained through a combination of 
our software framework and more advanced first principles approaches~\cite{GW,Hedin_GW_1965,Heyd2003,Liechtenstein1995,curtarolo:art86,curtarolo:art93,curtarolo:art103}.

In the same spirit as the N$(E)$ and E$_{gap}$, we consider the calculation of the ensemble average magnetic moment M of the system.
The magnetic moment of the \emph{i}th supercell is labeled as $M_i$.
If the ground state of the \emph{i}th structure is non-spin-polarized, then its magnetic moment is set to zero, \textit{i.e.}, $M_i=0$.
Taking into account the impact of signed spins on the ensemble average, this approach is limited only to ferromagnetic solutions.
Additionally, as an initialization for the self-consistent run, we assume the same ferromagnetic alignment among all of the spins in the system
(an AFLOW calculation standard)~\cite{AFLOW_STANDARD}.
Finally, the ensemble average magnetic moment of the system is calculated with the following formula:

\[\mathrm{M} = \sum\limits_{i = 1}^n {{P_i} \times } |{M_i}|.\]

\begin{figure*}[tb!]
\center
\includegraphics[width=\textwidth]{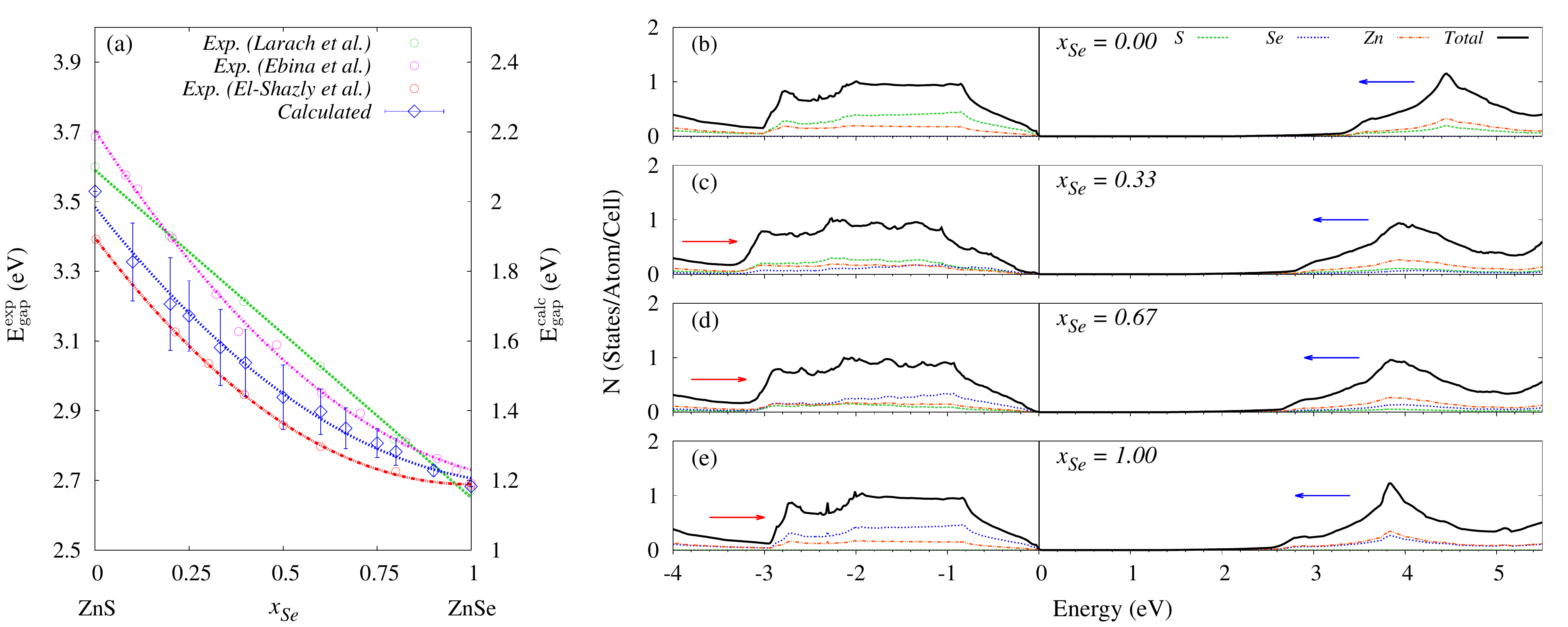}
{\caption{Disordered ZnS$_{1-x}$Se$_x$. (a) A comparison of the experimental \cite{Larach_PR_1957,Ebina_PRB_1974,El-Shazly_APA_1985} vs. 
calculated compositional dependence of the band gap energy E$_{gap}$ at room temperature. 
A rigid shift in the E${gap}$ axis relative to experimental results of ZnSe (second ordinate axis) accounts for the expected systematic 
deviation in DFT calculations~\cite{Perdew_IJQC_1985}.
Only the lowest empirical E$_{gap}$ trends are shown.
Error bars indicate the weighted standard deviation of the ensemble average E${gap}$.
(b)-(e) Calculated density of states plots for various compositions: 
(b) $x=0.00$ ($n=1$), 
(c) $x=0.33$ ($n=3$), 
(d) $x=0.67$ ($n=3$), and 
(e) $x=1.00$ ($n=1$).
The straight vertical line indicates the position of the valence band maximum. }}
\label{f1}
\end{figure*}

\section{Example applications}

To illustrate the effectiveness of our approach, we analyze disordered systems of technological importance:  
zinc chalcogenides, wide-gap oxide semiconductors, and iron alloys.
Unless otherwise stated, the supercells used in our calculations were generated with the lowest superlattice size $n_{xct}$ needed to represent
the composition exactly.

%\subsection{Disordered ZnS$_{1-x}$Se$_x$}
\subsection{Zinc chalcogenides}
Over the years, zinc chalcogenides have garnered interest for a dynamic range of applications---beginning with the creation 
of the first blue-light emitting laser diodes~\cite{Haase_APL_1991}, and recently have been studied 
as inorganic graphene analogues (IGAs) with potential applications in flexible and transparent nanodevices~\cite{Sun_NComm_2012}.
These wide-gap II-VI semiconductors have demonstrated a smoothly tunable band gap energy E$_{gap}$ with respect to composition~
\cite{Larach_PR_1957,Ebina_PRB_1974,El-Shazly_APA_1985}.
Both linear and quadratic dependencies have been observed, with the latter phenomenon referred to as \textit{optical bowing}~
\cite{Bernard_PRB_1986}.
Specifically, given the pseudo-ternary system A$_{x}$B$_{1-x}$C, 
\[
\mathrm{E}_{gap}(x)=\left[x \epsilon_{AC}+(1-x)\epsilon_{BC}\right] - b x(1-x),
\]
with $b$ characterizing the bowing.
While Larach \textit{et al.} reported a linear dependence ($b=0$)~\cite{Larach_PR_1957},
Ebina \textit{et al.} \cite{Ebina_PRB_1974} and 
El-Shazly \textit{et al.} \cite{El-Shazly_APA_1985} reported similar bowing parameters of 
$b=0.613\pm0.027$ eV and $b=0.457\pm0.044$ eV, respectively, averaged over the two observed direct transitions.

As a proof of concept, we utilized our developed disordered system framework to calculate the compositional dependence of the E$_{gap}$ and 
DOS for ZnS$_{1-x}$Se$_x$ at room temperature (annealed limit).
Overall, this system shows relatively low disorder ($\mathrm{max}\left(\Delta {H_{F,i}}\right)\sim 0.005$ eV), 
exhibiting negligible variations in the ensemble average properties at higher temperatures.
These results, illustrated in Fig. 1, are directly compared against experimental measurements~
\cite{Larach_PR_1957,Ebina_PRB_1974,El-Shazly_APA_1985}.
Common among all three trends (Fig. 1(a)) is the E$_{gap}$ shrinkage with increasing $x_{Se}$, 
as well as a near $1~\mathrm{eV}$ tunable E$_{gap}$ range.
Our calculated trend demonstrates a non-zero bowing similar to that observed by both Ebina \textit{et al.}~\cite{Ebina_PRB_1974} and 
El-Shazly \textit{et al.}~\cite{El-Shazly_APA_1985}.
A fit shows a bowing parameter of $b=0.585\pm0.078$ eV, lying in the range between the two experimental bowing parameters.

We also plot the ensemble average DOS plots at room temperature for $x=0.00$ ($n=1$), $0.33$ ($n=3$), $0.67$ ($n=3$), 
and $1.00$ ($n=1$) in Figs. 1(b)-1(e).
The plots echo the negatively correlated band gap relationship illustrated in Fig. 1(a), highlighting 
that the replacement of sulfur with selenium atoms reduces the band gap.
Specifically, we observe two phenomena as we increase the concentration of selenium:  (\textcolor{red}{red arrows}) 
the reduction of the valence band width
(with the exception of $x_{Se} = 0.00$ (ZnS) concentration), and (\textcolor{blue}{blue arrows}) 
a shift of the conduction band peak back towards the Fermi energy.
The valence band of ZnS more closely resembles that of its extreme concentration counterpart at $x_{Se} = 1.00$
(ZnSe) than the others.
The extreme concentration conduction peaks appear more defined than their intermediate concentration counterparts, which is likely an artifact
of the ensemble averaging calculation.

Finally, we consider a partial-DOS analysis in both species and orbitals (not shown).
In the valence band, sulfur and selenium account for the majority of the states, in agreement with their relative concentrations.
Meanwhile, zinc accounts for the majority of the states in the conduction band at all concentrations.
Correspondingly, at all concentrations, the $p$-orbitals make up the majority of the valence band, 
whereas the conduction band consists primarily of $s$- and $p$-orbitals.
These observations are consistent with conclusions drawn from previous optical reflectivity measurements that optical transitions
are possible from sulfur or selenium valence bands to zinc conduction bands~\cite{Kirschfeld_PRL_1972}.

Overall, our concentration-evolving E$_{gap}$ trend and DOS plots support a continuing line of work~
\cite{Larach_PR_1957,Ebina_PRB_1974,El-Shazly_APA_1985} corroborating that this system
is of the amalgamation type~\cite{Onodera_JPSJ_1968}
and not of the persistence type~\cite{Kirschfeld_PRL_1972}.
Notably, however, reflectivity spectra shows that the peak position in the E$_{gap}$ for ZnS rich alloys may remain stationary~
\cite{Ebina_PRB_1974},
which may have manifested itself in the aforementioned anomaly observed in this structure's valence band width.

\begin{figure*}[tb!]
\center
\includegraphics[width=\textwidth]{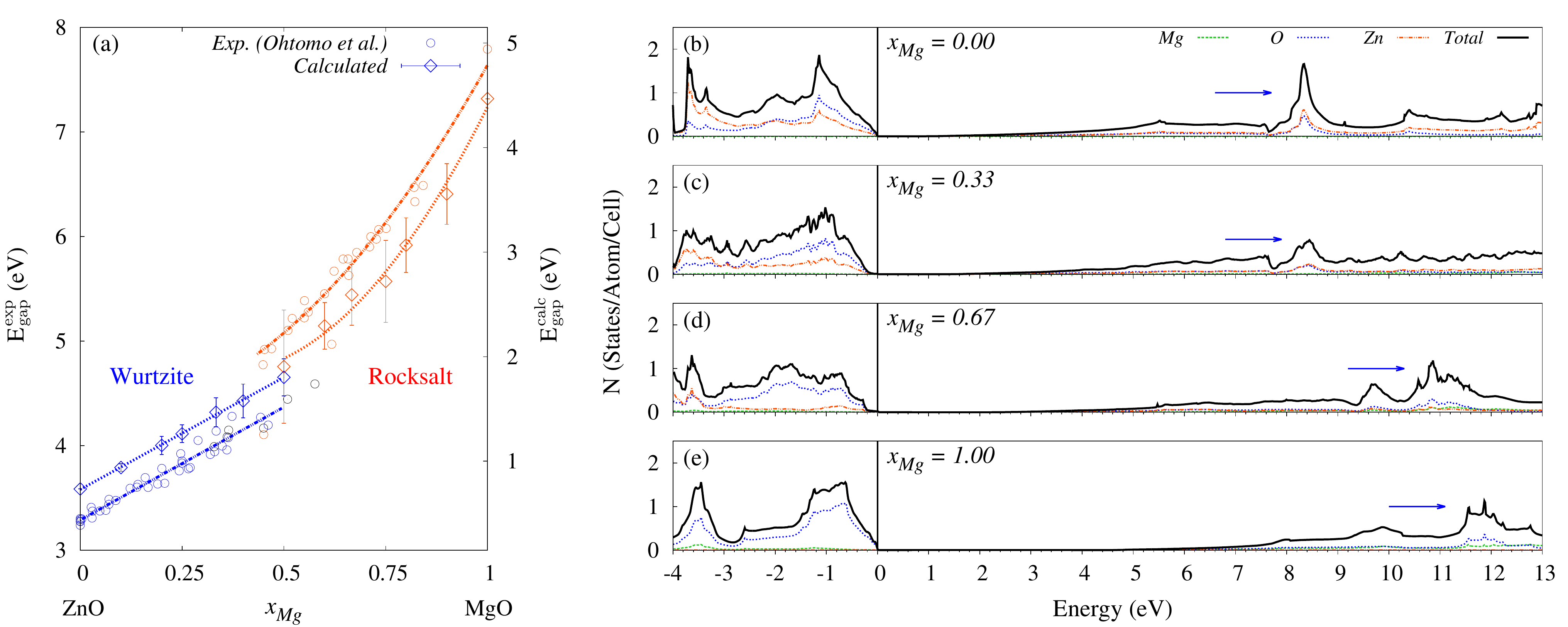}
{\caption{Disordered Mg$_{x}$Zn$_{1-x}$O. (a) A comparison of the experimental~\cite{Ohtomo_SST_2005,Takeuchi_JAP_2003,
Chen_JAPCM_2003,Takagi_JJAP_2003,Choopun_APL_2002,Minemoto_TSF_2000,Sharma_APL_1999,Ohtomo_APL_1998} vs. calculated compositional 
dependence of the band gap energy E$_{gap}$ at room temperature. 
A rigid shift in the E${gap}$ axis relative to experimental results of MgO (second ordinate axis) accounts for the expected 
systematic deviation in DFT calculations~\cite{Perdew_IJQC_1985}.
The wurtzite and rocksalt structures are highlighted in blue and red, respectively,
while the mixed phase structures are shown in black.
Error bars indicate the weighted standard deviation of the ensemble average E${gap}$.
(b)-(e) Calculated density of states plots for various compositions: 
(b) $x=0.00$ ($n=1$), 
(c) $x=0.33$ ($n=3$), 
(d) $x=0.67$ ($n=3$), and 
(e) $x=1.00$ ($n=1$). 
The straight vertical line indicates the position of the valence band maximum.}}
\label{f2}
\end{figure*}

\subsection{Wide-gap oxide semiconductor alloys}
Zinc oxide (ZnO) has proven to be a pervasive material, with far reaching applications such as paints, catalysts,
pharmaceuticals (sun creams), and optoelectronics~\cite{Takeuchi_MgZnO_Patent}.
It has long been investigated for its electronic properties, and falls into the class of transparent conducting 
oxides~\cite{Ellmer_ZnO_2007}.
Just as our previous zinc chalcogenide example, ZnO is a wide-gap II-VI semiconductor that has demonstrated
a smoothly tunable band gap energy E$_{gap}$ with composition.
In particular, ZnO has been engineered to have an E$_{gap}$ range as large as 5 eV by synthesizing it with magnesium. 
This pairing has been intensively studied because of the likeness in ionic radius between zinc and magnesium
which results in mitigated misfit strain in the heterostructure~\cite{Yoo_TSF_2015}.
While the solubility of MgO and ZnO is small, synthesis has been made possible throughout the full compositional
spectrum~\cite{Ohtomo_SST_2005,Takeuchi_JAP_2003,Chen_JAPCM_2003,Takagi_JJAP_2003,Choopun_APL_2002,
Minemoto_TSF_2000,Sharma_APL_1999,Ohtomo_APL_1998}.

As another proof of concept, we model the compositional dependence of the E$_{gap}$ and DOS for Mg$_{x}$Zn$_{1-x}$O at room temperature
(annealed limit).
In particular, we chose this disordered system to illustrate the breath of materials which this framework can model.
Similar to ZnS$_{1-x}$Se$_x$, this system shows relatively low disorder ($\mathrm{max}\left(\Delta {H_{F,i}}\right)\sim 0.007$ eV), 
exhibiting negligible variations in the ensemble average properties at higher temperatures.
We compare our results to that observed empirically~\cite{Ohtomo_SST_2005,Takeuchi_JAP_2003,Chen_JAPCM_2003,Takagi_JJAP_2003,Choopun_APL_2002,
Minemoto_TSF_2000,Sharma_APL_1999,Ohtomo_APL_1998} in Fig. 2.
As illustrated in Fig. 2(a), Ohtomo \textit{et al.} observed a composition dependent phase transition 
from a wurtzite to a rocksalt structure with increasing $x_{Mg}$; the transition occurring around the mid concentrations.
We mimic this transition in our calculations.
Empirically, the overall trend in the wurtzite phase shows a negligible bowing in the E$_{gap}$ trend, 
contrasting the significant bowing observed in the rocksalt phase.
We note that the wurtzite phase E$_{gap}$ trend shows a slope of $2.160\pm0.080$ eV, while the rocksalt phase shows a bowing
parameter of $3.591\pm0.856$ eV.
Calculated trends are shown in Fig. 2(a).
Qualitatively, we also observe linear and non-linear E$_{gap}$ trends in the wurtzite and rocksalt phases, respectively.
The fits were as follows:  we observe a slope of $2.147\pm0.030$ eV in the wurtzite phase and a bowing parameter of
$5.971\pm1.835$ eV in the rocksalt phase. 
These trends match experiment well within the margins of error.
We observe a larger margin of error in the rocksalt phase, particular in the phase separated region 
($0.4\lesssim x_{Mg} \lesssim 0.6$).
This may be indicative of the significant shear strain and complex nucleation behavior characterizing the region~\cite{Takeuchi_JAP_2003}.

We also plot the ensemble average DOS at room temperature for $x=0.00$ ($n=1$), $0.33$ ($n=3$), $0.67$ ($n=3$), and $1.00$ ($n=1$) 
in Figs. 2(b)-2(e).
The plots not only echo the positively correlated band gap relationship illustrated in Fig. 2(a), 
but exhibit the aforementioned change from a linear to non-linear trend.
This is most easily seen by observing the shift in the conduction band toward the Fermi energy, 
highlighted by the \textcolor{blue}{blue arrows}.
Contrasting ZnS$_{1-x}$Se$_x$, we do not observe a significant change in width of the valence band as we vary the stoichiometry.

Finally, we consider a partial-DOS analysis in both species and orbitals (not shown).
Overall, the constant oxygen backbone plays a major role in defining the shape of both the valence and conduction bands,
particularly as $x_{Mg}$ increases.
This resonates with the strong $p$-orbital presence in both bands throughout all concentrations.
Zinc and its $d$-orbitals play a particularly dominant role in the valence band in magnesium-poor structures.

\begin{figure*}[ht!]
\center
\includegraphics[width=\textwidth]{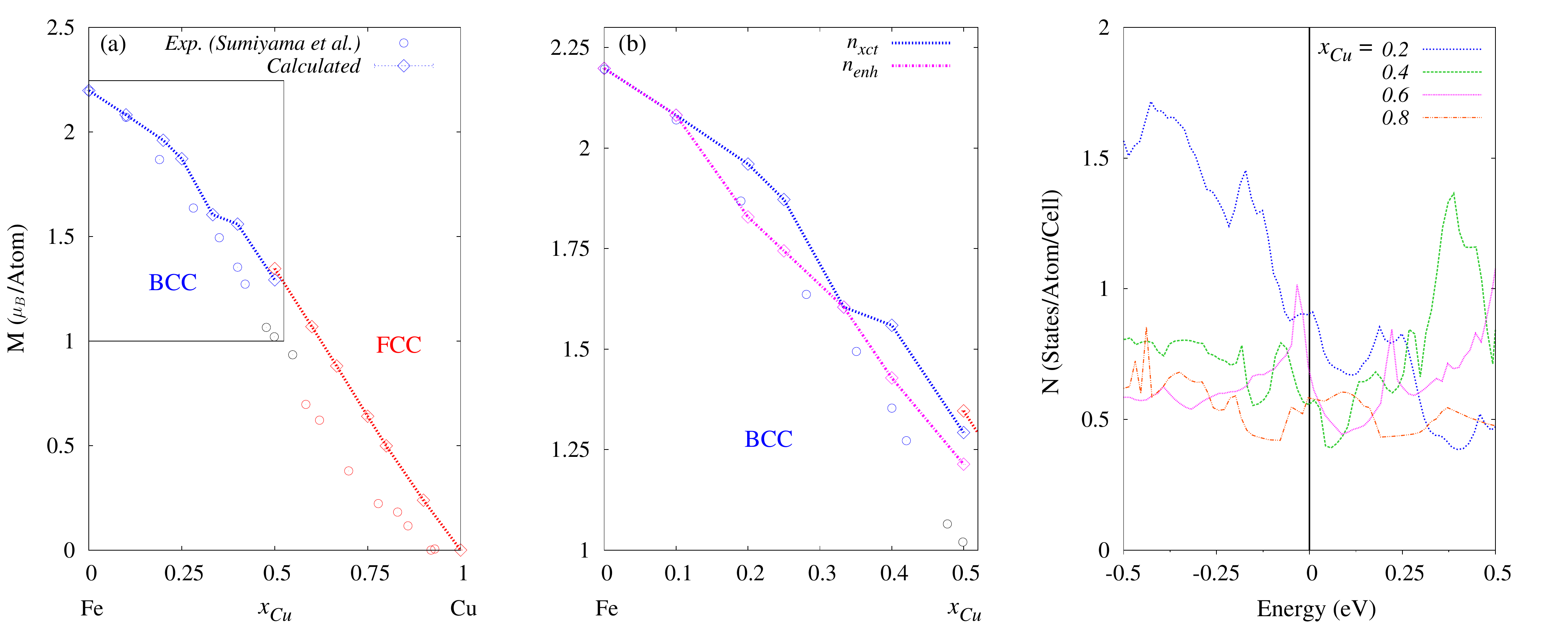}
{\caption{ Disordered Fe$_{1-x}$Cu$_{x}$. (a) A comparison of the experimental~\cite{Sumiyama_JPSJ_1984} vs. calculated compositional 
dependence of the magnetic moment M. 
Our calculations mimic the following observed phases at 4.2 K: 
$x\leq0.42$ BCC phase,
$0.42<x<0.58$ mixed BCC-FCC phases,
$x\geq0.58$ FCC phase.
Error bars indicate the weighted standard deviation of the ensemble average E${gap}$.
(b) A comparison of the aforementioned trends with calculations performed with enhanced superlattice sizes $n$.
(c) Calculated unpolarized density of states (DOS) plots at $x_{Cu}=0.2$ ($n=5$), $0.4$ ($n=5$), $0.6$ ($n=5$), $0.8$ ($n=5$).}}
\label{f3}
\end{figure*}

\subsection{Iron alloys}
Despite its ubiquity, iron remains at the focus of critical materials research.
Even as new phenomena are discovered with an ever-growing effort to explore extreme conditions~
\cite{Bonetti_PRL_1999,Shimizu_Nature_2001,Bose_PRB_2003},
there exist long-standing, interesting aspects that are not fully resolved.
This includes the magnetic character of the (fcc) $\gamma$-Fe phase at low temperatures~
\cite{Gorria_PRB_2004,Abrahams_PR_1962,Sandratskii_AP_1998}, among other complexities in its magnetic phase diagram~\cite{Pepperhoff_Fe_2013}.
One popular approach to studying the $\gamma$-Fe phase is through the Fe$_{1-x}$Cu$_{x}$ disordered alloy~
\cite{Gorria_PRB_2004,Gorria_PRB_2005,Orecchini_JAC_2006}.
Nominally, unary copper and iron metals with fcc structures are nonmagnetic, but together exhibit ferromagnetic ordering with very 
high magnetic moments.
This observation has led to identification of Invar and anti-Invar behaviors, which may pave the way to enhanced 
thermomechanical actuators~\cite{Gorria_PRB_2004,Gorria_PRB_2005}.
Fe$_{1-x}$Cu$_{x}$ is an interesting structure in its own right, as it has extremely low miscibility~\cite{Liu_PRB_2005}.
Overcoming the hurdle of developing metastable structures throughout the full compositional range has been the focus of much
research~\cite{Ma_PMS_2005}.
Such metastable structures have demonstrated novel properties like high thermal and electrical conductivity~\cite{Korn_ZPBCM_1976},
magnetoresistance, and coercivity~\cite{Hihara_JJap_1997}.

As a final proof of concept, we model the compositional dependence of 
the magnetic moment M for Fe$_{1-x}$Cu$_{x}$ at $T=4.2$ K for direct comparison against experimental results~\cite{Sumiyama_JPSJ_1984}.
Considering both the sensitivity of magnetic properties to temperature as well as the significant disorder exhibited in 
this system ($\mathrm{max}\left(\Delta {H_{F,i}}\right)\sim 1.63$ eV), we limit our analysis to the low temperature limit.
This is also where we expect our framework to perform optimally, which considers structures relaxed at zero temperature and pressure
~\cite{AFLOW_STANDARD}.
The results are illustrated in Fig. 3.
Sumiyama \textit{et al.} shows that the disordered system's phase is concentration dependent, 
with a phase transition from bcc to fcc in the mid concentrations as $x_{Cu}$ increases.
Just as with Mg$_{x}$Zn$_{1-x}$O, we mimic these observations in our calculation.
The overall decreasing trend in M with reduced $x_{Fe}$ in Fig. 3(a) matches our expectations well.

With such a simple system, we also explored whether an augmented superlattice size $n$ enhances our results.
While the concentration remains constant for $n$ above that which is needed for the desired concentration $n_{xct}$, 
more structures are introduced into the ensemble average.
The structures themselves also increase in size by a factor of $n$ relative to their parent structure.
For $x_{Cu}=0.2,~0.4$, we simply doubled $n$ ($n_{enh}=10$), while we tripled $n$ for $x_{Cu}=0.25$ ($n_{enh}=12$).
With only two two-atom structures needed to describe $x_{Cu}=0.5$ at $n_{xct}$, we were able to increase $n$ by a factor of five ($n_{enh}=10$) without
compromising the feasibility of the calculation.
A comparison of results calculated at $n_{enh}$ is shown in Fig. 3(b).
At most concentrations, we observed substantial improvements as our calculated trend more closely follows that which was observed 
empirically.

Finally, we consider this system's ensemble average DOS in Fig. 3(c).
In general, the DOS near the Fermi energy decreases with increasing $x_{Cu}$, with some instability near the mixed phase regions.
This can be understood using the Stoner criterion model for transitional metals~\cite{Xie_CMS_2011,James_PRB_1999}.
Namely, ferromagnetism appears when the gain in exchange energy is larger than the loss in kinetic energy.
A larger DOS at the Fermi energy induces a higher exchange energy and favors a split into the ferromagnetic state.
The competition between ferromagnetic and paramagnetic phases can be inferred from the decreasing M
trend as depicted in Fig. 3(a).

\section{Additional Materials}
The automated approach to model disordered systems discussed here is implemented with the AFLOW framework~\cite{AFLOW_STANDARD,aflowPAPER}
and can be downloaded from \href{http://aflowlib.org}{\texttt{http://aflowlib.org}}.

\section{Conclusion}
In this work, we have introduced a software framework capable of modeling substitutionally disordered materials.
Specifically, the framework delivers high value properties of disordered systems, including the density of states (DOS), 
band gap energy E$_{gap}$, and magnetic moment M.
Though a number of technologically significant examples, we have illustrated the prowess of this highly efficient and 
convenient framework.
Such materials that exhibit highly tunable properties are of critical importance toward the goal of rational materials design.
Without loss of feasibility or accuracy, the framework exploits highly successful high-throughput first principles approaches 
in more complex, real-world systems.

\begin{acknowledgments}
We thank C. Toher, M. Buongiorno Nardelli, and M. Fornari for various
technical discussions that have contributed to the results reported in
this article.
This work was supported by ONR-MURI under Contract N00014-13-1-0635, DOD
ONR (N00014-14-1-0526), and the Duke University Center for Materials
Genomics.
K. Yang acknowledges the support by start-up funds from the University of California San Diego.
C. Oses acknowledges support from the National Science Foundation Graduate Research Fellowship under Grant No. DGF1106401.
We also acknowledge the CRAY corporation for computational support.
\end{acknowledgments}

\bibliographystyle{PhysRevwithTitles_noDOI_v1b}

\begin{thebibliography}{10}
\expandafter\ifx\csname urlstyle\endcsname\relax
  \providecommand{\doi}[1]{doi:\discretionary{}{}{}#1}\else
  \providecommand{\doi}{doi:\discretionary{}{}{}\begingroup
  \urlstyle{rm}\Url}\fi
\providecommand{\selectlanguage}[1]{\relax}
\providecommand{\bibAnnoteFile}[1]{%
  \IfFileExists{#1}{\begin{quotation}\noindent\textsc{Key:} #1\\
  \textsc{Annotation:}\ \input{#1}\end{quotation}}{}}
\providecommand{\bibAnnote}[2]{%
  \begin{quotation}\noindent\textsc{Key:} #1\\
  \textsc{Annotation:}\ #2\end{quotation}}

\bibitem{DFT}
W.~Kohn and L.~J. Sham, \emph{Self-consistent equations including exchange and
  correlation effects}, Phys.\ Rev. \textbf{140}, A1133 (1965).
\input{DFT}

\bibitem{Hohenberg_PR_1964}
P.~Hohenberg and W.~Kohn, \emph{Inhomogeneous Electron Gas}, Phys.\ Rev.
  \textbf{136}, B864--B871 (1964).
\input{Hohenberg_PR_1964}

\bibitem{Elliott_PoAM_1990}
S.~R. Elliott, \emph{Physics of Amorphous Materials} (Longman Scientific \&
  Technical, 1990).
\input{Elliott_PoAM_1990}

\bibitem{Xie_ACB_2015}
L.~Xie, P.~Brault, C.~Coutanceau, J.~Bauchire, A.~Caillard, S.~Baranton,
  J.~Berndt, and E.~C. Neyts, \emph{Efficient amorphous platinum catalyst
  cluster growth on porous carbon: A combined molecular dynamics and
  experimental study}, Appl.\ Catal.\ B:\ Environ. \textbf{162}, 21--26 (2015).
\input{Xie_ACB_2015}

\bibitem{Kurian_JPCC_2013}
S.~Kurian, H.~Seo, and H.~Jeon, \emph{Significant Enhancement in Visible Light
  Absorption of {TiO$_2$} Nanotube Arrays by Surface Band Gap Tuning}, J.\
  Phys.\ Chem.\ C \textbf{117}, 16811--16819 (2013).
\input{Kurian_JPCC_2013}

\bibitem{Bednorz_ZPBCM_1986}
J.~G. Bednorz and K.~A. M\"{u}ller, \emph{Possible high {T$_c$}
  superconductivity in the {B}a${−-}${L}a${-}${C}u${-}${O} system}, Z.\
  Phys.\ B\ Con.\ Mat. \textbf{64}, 189--193 (1986).
\input{Bednorz_ZPBCM_1986}

\bibitem{Maeno_Nature_1994}
Y.~Maeno, H.~Hashimoto, K.~Yoshida, S.~Nishizaki, T.~Fujita, J.~G. Bednorz, and
  F.~Lichtenberg, \emph{Superconductivity in a layered perovskite without
  copper}, Nature \textbf{372}, 532--534 (1994).
\input{Maeno_Nature_1994}

\bibitem{Winter_JACerS_2007}
M.~R. Winter and D.~R. Clarke, \emph{Oxide Materials with Low Thermal
  Conductivity}, J.\ Am.\ Ceram Soc. \textbf{90}, 533--540 (2007).
\input{Winter_JACerS_2007}

\bibitem{Patra_JAP_2012}
N.~C. Patra, S.~Bharatan, J.~Li, M.~Tilton, and S.~Iyer, \emph{Molecular beam
  epitaxial growth and characterization of {InSb$_{1−-x}$N$_{x}$} on {GaAs}
  for long wavelength infrared applications}, J.\ Appl.\ Phys. \textbf{111}
  (2012).
\input{Patra_JAP_2012}

\bibitem{Wang_SR_2013}
H.~Wang, Y.~N. Zhang, R.~Q. Wu, L.~Z. Sun, D.~S. Xu, and Z.~D. Zhang,
  \emph{Understanding strong magnetostriction in {Fe$_{100−-x}$Ga$_{x}$}
  alloys}, Sci.\ Rep. \textbf{3} (2013).
\input{Wang_SR_2013}

\bibitem{Nord_1931_AP_VCA}
L.~Nordheim, \emph{Zur Elektronentheorie der Metalle {I}}, Ann. Phys. (Leipzig)
  \textbf{9}, 607--640 (1931).
\input{Nord_1931_AP_VCA}

\bibitem{Vanderbilt_2000_PRB_VCA}
L.~Bellaiche and D.~Vanderbilt, \emph{Virtual crystal approximation revisited:
  Application to dielectric and piezoelectric properties of perovskites},
  Phys.\ Rev.\ B \textbf{61}, 7877--7882 (2000).
\input{Vanderbilt_2000_PRB_VCA}

\bibitem{Soven_PhysRev_1967}
P.~Soven, \emph{Coherent-Potential Model of Substitutional Disordered Alloys},
  Phys.\ Rev. \textbf{156}, 809--813 (1967).
\input{Soven_PhysRev_1967}

\bibitem{Korringa1947392}
J.~Korringa, \emph{On the calculation of the energy of a Bloch wave in a
  metal}, Physica \textbf{13}, 392--400 (1947).
\input{Korringa1947392}

\bibitem{Kohn_1954_PhysRev}
W.~Kohn and N.~Rostoker, \emph{Solution of the {Schr\"{o}dinger} Equation in
  Periodic Lattices with an Application to Metallic Lithium}, Phys.\ Rev.
  \textbf{94}, 1111--1120 (1954).
\input{Kohn_1954_PhysRev}

\bibitem{Stocks_PRL_1978}
G.~M. Stocks, W.~M. Temmerman, and B.~L. Gyorffy, \emph{Complete Solution of
  the {Korringa-Kohn-Rostoker Coherent-Potential-Approximation} Equations:
  {Cu-Ni} Alloys}, Phys.\ Rev.\ Lett. \textbf{41}, 339--343 (1978).
\input{Stocks_PRL_1978}

\bibitem{Faulkner_PM_2006}
J.~S. Faulkner, S.~Pella, A.~Rusanu, Y.~Puzyrev, T.~Leventouri, G.~M. Stocks,
  and B.~Ujfalussy, \emph{Mean-field approximations for the electronic states
  in disordered alloys}, Philos.\ Mag. \textbf{86}, 2661--2671 (2006).
\input{Faulkner_PM_2006}

\bibitem{Faulkner_PMS_1982}
J.~S. Faulkner, \emph{The modern theory of alloys}, Prog.\ Mater.\ Sci.
  \textbf{27}, 1--187 (1982).
\input{Faulkner_PMS_1982}

\bibitem{Zunger_PRL_1990}
A.~Zunger, S.-H. Wei, L.~G. Ferreira, and J.~E. Bernard, \emph{Special
  quasirandom structures}, Phys.\ Rev.\ Lett. \textbf{65}, 353--356 (1990).
\input{Zunger_PRL_1990}

\bibitem{Shan_PRL_1999}
W.~Shan, W.~Walukiewicz, J.~W. Ager, E.~E. Haller, J.~F. Geisz, D.~J. Friedman,
  J.~M. Olson, and S.~R. Kurtz, \emph{Band Anticrossing in {GaInNAs} Alloys},
  Phys.\ Rev.\ Lett. \textbf{82}, 1221--1224 (1999).
\input{Shan_PRL_1999}

\bibitem{Popescu_PRL_2010}
V.~Popescu and A.~Zunger, \emph{Effective Band Structure of Random Alloys},
  Phys.\ Rev.\ Lett. \textbf{104}, 236403 (2010).
\input{Popescu_PRL_2010}

\bibitem{Habgood_PCCP_2011}
M.~Habgood, R.~Grau-Crespo, and S.~L. Price, \emph{Substitutional and
  orientational disorder in organic crystals: a symmetry-adapted ensemble
  model}, Phys.\ Chem.\ Chem.\ Phys. \textbf{13}, 9590--9600 (2011).
\input{Habgood_PCCP_2011}

\bibitem{Haverkort_ArXiv_2011}
M.~W. Haverkort, I.~S. Elfimov, and G.~A. {Sawatzky}, \emph{Electronic
  structure and self energies of randomly substituted solids using density
  functional theory and model calculations}, arXiv:1109.4036v1
  [cond-mat.mtrl-sci]  (2011).
\input{Haverkort_ArXiv_2011}

\bibitem{nmatHT}
S.~Curtarolo, G.~L.~W. Hart, M.~{Buongiorno~Nardelli}, N.~Mingo, S.~Sanvito,
  and O.~Levy, \emph{The {high-throughput} highway to computational materials
  design}, Nat.\ Mater. \textbf{12}, 191--201 (2013).
\input{nmatHT}

\bibitem{MGI_OSTP}
{Office of Science and Technology Policy, White House}, \emph{Materials Genome
  Initiative for Global Competitiveness}, \\{\sf http://www.whitehouse.gov/mgi}
   (2011).
\input{MGI_OSTP}

\bibitem{AFLOW_STANDARD}
C.~E. Calderon, J.~J. Plata, C.~Toher, C.~Oses, O.~Levy, M.~Fornari, A.~Natan,
  M.~J. Mehl, G.~Hart, M.~{Buongiorno Nardelli}, and S.~Curtarolo, \emph{The
  {AFLOW} standard for high-throughput materials science calculations}, Comp.\
  Mat.\ Sci. \textbf{108, Part A}, 233--238 (2015).
\input{AFLOW_STANDARD}

\bibitem{aflowPAPER}
S.~Curtarolo, W.~Setyawan, G.~L.~W. Hart, M.~Jahnatek, R.~V. Chepulskii, R.~H.
  Taylor, S.~Wang, J.~Xue, K.~Yang, O.~Levy, M.~Mehl, H.~T. Stokes, D.~O.
  Demchenko, , and D.~Morgan, \emph{{AFLOW}: an automatic framework for
  high-throughput materials discovery}, Comp.\ Mat.\ Sci. \textbf{58}, 218--226
  (2012).
\input{aflowPAPER}

\bibitem{Rappe_1992_JCAS_UFF}
A.~K. Rappe, C.~J. Casewit, K.~S. Colwell, W.~A. Goddard, and W.~M. Skiff,
  \emph{{UFF}, a full periodic table force field for molecular mechanics and
  molecular dynamics simulations}, J.\ Am.\ Chem.\ Soc. \textbf{114},
  10024--10035 (1992).
\input{Rappe_1992_JCAS_UFF}

\bibitem{Setyawan_2010_CMS.299}
W.~Setyawan and S.~Curtarolo, \emph{High-throughput electronic band structure
  calculations: Challenges and tools}, Comp.\ Mat.\ Sci. \textbf{49}, 299--312
  (2010).
\input{Setyawan_2010_CMS.299}

\bibitem{VASP_PRB}
G.~Kresse and J.~Furthm\"uller, \emph{Efficient iterative schemes for {\it ab
  initio} total-energy calculations using a plane-wave basis set}, Phys.\ Rev.\
  B \textbf{54}, 11169--11186 (1996).
\input{VASP_PRB}

\bibitem{Gus_2008_PRB_ENUM}
G.~L.~W. Hart and R.~W. Forcade, \emph{Algorithm for generating derivative
  structures}, Phys.\ Rev.\ B \textbf{77}, 224115 (2008).
\input{Gus_2008_PRB_ENUM}

\bibitem{Perdew_IJQC_1985}
J.~P. Perdew, \emph{Density functional theory and the band gap problem}, Int.\
  J.\ Quantum.\ Chem. \textbf{28}, 497--523 (1985).
\input{Perdew_IJQC_1985}

\bibitem{GW}
F.~Aryasetiawan and O.~Gunnarsson, \emph{The {GW} Method}, Rep. Prog. Phys.
  \textbf{61}, 237 (1998).
\input{GW}

\bibitem{Hedin_GW_1965}
L.~Hedin, \emph{New Method for Calculating the One-Particle Green's Function
  with Application to the Electron-Gas Problem}, Phys. Rev. \textbf{139},
  A796--A823 (1965).
\input{Hedin_GW_1965}

\bibitem{Heyd2003}
J.~Heyd, G.~E. Scuseria, and M.~Ernzerhof, \emph{Hybrid functionals based on a
  screened Coulomb potential}, J. Chem. Phys. \textbf{118}, 8207--8215 (2003).
\input{Heyd2003}

\bibitem{Liechtenstein1995}
A.~I. Liechtenstein, V.~I. Anisimov, and J.~Zaanen, \emph{{Density-functional}
  theory and strong interactions: Orbital ordering in {Mott-Hubbard}
  insulators}, Phys. Rev. B \textbf{52}, R5467--R5470 (1995).
\input{Liechtenstein1995}

\bibitem{curtarolo:art86}
L.~A. Agapito, A.~Ferretti, A.~Calzolari, S.~Curtarolo, and
  M.~{Buongiorno~Nardelli}, \emph{Effective and accurate representation of
  extended Bloch states on finite Hilbert spaces}, Phys.\ Rev.\ B \textbf{88},
  165127 (2013).
\input{curtarolo:art86}

\bibitem{curtarolo:art93}
L.~A. Agapito, S.~Curtarolo, and M.~B. Nardelli, \emph{Reformulation of DFT+$U$
  as a pseudo-hybrid {Hubbard} density functional}, Phys.\ Rev.\ X \textbf{5},
  011006 (2015).
\input{curtarolo:art93}

\bibitem{curtarolo:art103}
P.~Gopal, M.~Fornari, S.~Curtarolo, L.~A. Agapito, L.~Liyanage, and
  M.~{Buongiorno Nardelli}, \emph{Improved predictions of the physical
  properties of {Zn}- and {Cd}-based wide band-gap semiconductors: a validation
  of the {ACBN0} functional}, Phys.\ Rev.\ B \textbf{91}, 245202 (2015).
\input{curtarolo:art103}

\bibitem{Larach_PR_1957}
S.~Larach, R.~E. Shrader, and C.~F. Stocker, \emph{Anomalous Variation of Band
  Gap with Composition in {Zinc Sulfo- and Seleno-Tellurides}}, Phys.\ Rev.
  \textbf{108}, 587--589 (1957).
\input{Larach_PR_1957}

\bibitem{Ebina_PRB_1974}
A.~Ebina, E.~Fukunaga, and T.~Takahashi, \emph{Variation with composition of
  the ${E}_{0}$ and ${E}_{0}+{\ensuremath{\Delta}}_{0}$ gaps in
  {$ZnS_{x}Se_{1-x}$} alloys}, Phys.\ Rev.\ B \textbf{10}, 2495--2500 (1974).
\input{Ebina_PRB_1974}

\bibitem{El-Shazly_APA_1985}
A.~A. El-Shazly, M.~M.~H. El-Naby, M.~A. Kenawy, M.~M. El-Nahass, H.~T.
  El-Shair, and A.~M. Ebrahim, \emph{Optical properties of ternary
  {ZnS$_{x}$Se$_{1-−x}$} polycrystalline thin films}, Appl.\ Phys.\ A\
  Mater.\ Sci.\ Process. \textbf{36}, 51--53 (1985).
\input{El-Shazly_APA_1985}

\bibitem{Haase_APL_1991}
M.~A. Haase, J.~Qiu, J.~M. DePuydt, and H.~Cheng, \emph{{Blue-green} laser
  diodes}, Appl.\ Phys.\ Lett. \textbf{59}, 1272--1274 (1991).
\input{Haase_APL_1991}

\bibitem{Sun_NComm_2012}
Y.~Sun, Z.~Sun, S.~Gao, H.~Cheng, Q.~Liu, J.~Piao, T.~Yao, C.~Wu, S.~Hu,
  S.~Wei, and Y.~Xie, \emph{Fabrication of flexible and freestanding zinc
  chalcogenide single layers}, Nat.\ Commun. \textbf{3}, 1057 (2012).
\input{Sun_NComm_2012}

\bibitem{Bernard_PRB_1986}
J.~E. Bernard and A.~Zunger, \emph{Optical bowing in zinc chalcogenide
  semiconductor alloys}, Phys.\ Rev.\ B \textbf{34}, 5992--5995 (1986).
\input{Bernard_PRB_1986}

\bibitem{Kirschfeld_PRL_1972}
K.~E. Kirschfeld, N.~Nelkowski, and T.~S. Wagner, \emph{Optical Reflectivity
  and Band Structure of {$ZnS_{1-x}Se_{x}$} Mixed Crystals}, Phys.\ Rev.\ Lett.
  \textbf{29}, 66--68 (1972).
\input{Kirschfeld_PRL_1972}

\bibitem{Onodera_JPSJ_1968}
Y.~Onodera and Y.~Toyozawa, \emph{Persistence and Amalgamation Types in the
  Electronic Structure of Mixed Crystals}, J.\ Phys.\ Soc.\ Jpn. \textbf{24},
  341--355 (1968).
\input{Onodera_JPSJ_1968}

\bibitem{Ohtomo_SST_2005}
A.~Ohtomo and A.~Tsukazaki, \emph{Pulsed laser deposition of thin films and
  superlattices based on {ZnO}}, Semicond.\ Sci.\ Tech. \textbf{20}, S1 (2005).
\input{Ohtomo_SST_2005}

\bibitem{Takeuchi_JAP_2003}
I.~Takeuchi, W.~Yang, K.-S. Chang, M.~A. Aronova, T.~Venkatesan, R.~D. Vispute,
  and L.~A. Bendersky, \emph{Monolithic multichannel ultraviolet detector
  arrays and continuous phase evolution in {Mg$_{x}$Zn$_{1-−x}$O} composition
  spreads}, Journal of Applied Physics \textbf{94}, 7336--7340 (2003).
\input{Takeuchi_JAP_2003}

\bibitem{Chen_JAPCM_2003}
J.~Chen, W.~Z. Shen, N.~B. Chen, D.~J. Qiu, and H.~Z. Wu, \emph{The study of
  composition non-uniformity in ternary {Mg$_{x}$Zn$_{1-−x}$O} thin films},
  J.\ Phys.:\ Conden.\ Matt. \textbf{15}, L475 (2003).
\input{Chen_JAPCM_2003}

\bibitem{Takagi_JJAP_2003}
T.~Takagi, H.~Tanaka, S.~Fujita, and S.~Fujita, \emph{Molecular Beam Epitaxy of
  High Magnesium Content Single-Phase Wurzite {Mg$_{x}$Zn$_{1-−x}$O} Alloys
  ($x \simeq 0.5$) and Their Application to Solar-Blind Region Photodetectors},
  Jpn.\ J.\ Appl.\ Phys \textbf{42}, L401 (2003).
\input{Takagi_JJAP_2003}

\bibitem{Choopun_APL_2002}
S.~Choopun, R.~D. Vispute, W.~Yang, R.~P. Sharma, T.~Venkatesan, and H.~Shen,
  \emph{Realization of band gap above 5.0 {eV} in metastable cubic-phase
  {Mg$_{x}$Zn$_{1-−x}$O} alloy films}, Appl.\ Phys.\ Lett. \textbf{80},
  1529--1531 (2002).
\input{Choopun_APL_2002}

\bibitem{Minemoto_TSF_2000}
T.~Minemoto, T.~Negami, S.~Nishiwaki, H.~Takakura, and Y.~Hamakawa,
  \emph{Preparation of {Zn$_{1-−x}$Mg$_{x}$O} films by radio frequency
  magnetron sputtering}, {Thin Solid Films} \textbf{372}, 173--176 (2000).
\input{Minemoto_TSF_2000}

\bibitem{Sharma_APL_1999}
A.~K. Sharma, J.~Narayan, J.~F. Muth, C.~W. Teng, C.~Jin, A.~Kvit, R.~M.
  Kolbas, and O.~W. Holland, \emph{Optical and structural properties of
  epitaxial {Mg$_{x}$Zn$_{1-−x}$O} alloys}, Appl.\ Phys.\ Lett. \textbf{75},
  3327--3329 (1999).
\input{Sharma_APL_1999}

\bibitem{Ohtomo_APL_1998}
A.~Ohtomo, M.~Kawasaki, T.~Koida, K.~Masubuchi, H.~Koinuma, Y.~Sakurai,
  Y.~Yoshida, T.~Yasuda, and Y.~Segawa, \emph{{Mg$_{x}$Zn$_{1-−x}$O} as a
  {II-VI} widegap semiconductor alloy}, Appl.\ Phys.\ Lett. \textbf{72},
  2466--2468 (1998).
\input{Ohtomo_APL_1998}

\bibitem{Takeuchi_MgZnO_Patent}
I.~Takeuchi, W.~Yang, K.~S. Chang, R.~D. Vispute, and T.~V. Venkatesan,
  \emph{System and method of fabrication and application of thin-films with
  continuously graded or discrete physical property parameters to functionally
  broadband monolithic microelectronic optoelectronic/sensor/actuator device
  arrays} (2007). US Patent 7,309,644.
\input{Takeuchi_MgZnO_Patent}

\bibitem{Ellmer_ZnO_2007}
K.~Ellmer, A.~Klein, and B.~Rech, \emph{Transparent Conductive Zinc Oxide:
  Basics and Applications in Thin Film Solar Cells}, Springer Series in
  Materials Science (Springer Berlin Heidelberg, 2007).
\input{Ellmer_ZnO_2007}

\bibitem{Yoo_TSF_2015}
S.~J. Yoo, J.-H. Lee, C.-Y. Kim, C.~H. Kim, J.~W. Shin, H.~S. Kim, and J.-G.
  Kim, \emph{Direct observation of the crystal structure changes in the
  {Mg$_{x}$Zn$_{1-−x}$O} alloy system}, {Thin Solid Films} \textbf{588},
  50--55 (2015).
\input{Yoo_TSF_2015}

\bibitem{Sumiyama_JPSJ_1984}
K.~Sumiyama, T.~Yoshitake, and Y.~Nakamura, \emph{Magnetic Properties of
  Metastable bcc and fcc {Fe-Cu} Alloys Produced by Vapor Quenching}, J.\
  Phys.\ Soc.\ Jpn. \textbf{53}, 3160--3165 (1984).
\input{Sumiyama_JPSJ_1984}

\bibitem{Bonetti_PRL_1999}
E.~Bonetti, L.~D. Bianco, D.~Fiorani, D.~Rinaldi, R.~Caciuffo, and A.~Hernando,
  \emph{Disordered Magnetism at the Grain Boundary of Pure Nanocrystalline
  Iron}, Phys.\ Rev.\ Lett. \textbf{83}, 2829--2832 (1999).
\input{Bonetti_PRL_1999}

\bibitem{Shimizu_Nature_2001}
K.~Shimizu, T.~Kimura, S.~Furomoto, K.~Takeda, K.~Kontani, Y.~Onuki, and
  K.~Amaya, \emph{Superconductivity in the non-magnetic state of iron under
  pressure}, Nature \textbf{412}, 316--318 (2001).
\input{Shimizu_Nature_2001}

\bibitem{Bose_PRB_2003}
S.~K. Bose, O.~V. Dolgov, J.~Kortus, O.~Jepsen, and O.~K. Andersen,
  \emph{Pressure dependence of electron-phonon coupling and superconductivity
  in hcp {Fe}: A linear response study}, Phys.\ Rev.\ B \textbf{67}, 214518
  (2003).
\input{Bose_PRB_2003}

\bibitem{Gorria_PRB_2004}
P.~Gorria, D.~Mart\'{i}nez-Blanco, J.~A. Blanco, A.~Hernando, J.~S.
  Garitaonandia, L.~F. Barqu\'{i}n, J.~Campo, and R.~I. Smith, \emph{Invar
  effect in fcc-{FeCu} solid solutions}, Phys.\ Rev.\ B \textbf{69}, 214421
  (2004).
\input{Gorria_PRB_2004}

\bibitem{Abrahams_PR_1962}
S.~C. Abrahams, L.~Guttman, and J.~S. Kasper, \emph{Neutron Diffraction
  Determination of Antiferromagnetism in Face-Centered Cubic
  ($\ensuremath{\gamma}$) Iron}, Phys.\ Rev. \textbf{127}, 2052--2055 (1962).
\input{Abrahams_PR_1962}

\bibitem{Sandratskii_AP_1998}
L.~M. Sandratskii, \emph{Noncollinear magnetism in itinerant-electron systems:
  Theory and applications}, Adv.\ Phys. \textbf{47}, 91--160 (1998).
\input{Sandratskii_AP_1998}

\bibitem{Pepperhoff_Fe_2013}
W.~Pepperhoff and M.~Acet, \emph{Constitution and Magnetism of Iron and its
  Alloys}, Engineering Materials (Springer Berlin Heidelberg, 2001).
\input{Pepperhoff_Fe_2013}

\bibitem{Gorria_PRB_2005}
P.~Gorria, D.~Mart\'{i}nez-Blanco, J.~A. Blanco, M.~J. P\'erez, A.~Hernando,
  L.~F. Barqu\'{i}n, and R.~I. Smith, \emph{High-temperature induced
  ferromagnetism on $\ensuremath{\gamma}\text{-}{Fe}$ precipitates in {FeCu}
  solid solutions}, Phys.\ Rev.\ B \textbf{72}, 014401 (2005).
\input{Gorria_PRB_2005}

\bibitem{Orecchini_JAC_2006}
A.~Orecchini, F.~Sacchetti, C.~Petrillo, P.~Postorino, A.~Congeduti,
  C.~Giorgetti, F.~Baudelet, and G.~Mazzone, \emph{Magnetic states of iron in
  metastable fcc {Fe}-{Cu} alloys}, J.\ Alloys Compound. \textbf{424}, 27--32
  (2006).
\input{Orecchini_JAC_2006}

\bibitem{Liu_PRB_2005}
J.~Z. Liu, A.~{van de Walle}, G.~Ghosh, and M.~Asta, \emph{Structure,
  energetics, and mechanical stability of {Fe-Cu} bcc alloys from
  first-principles calculations}, Phys.\ Rev.\ B \textbf{72}, 144109 (2005).
\input{Liu_PRB_2005}

\bibitem{Ma_PMS_2005}
E.~Ma, \emph{Alloys created between immiscible elements}, Prog.\ Mater.\ Sci.
  \textbf{50}, 413--509 (2005).
\input{Ma_PMS_2005}

\bibitem{Korn_ZPBCM_1976}
D.~Korn, H.~Pfeifle, and J.~Niebuhr, \emph{Electrical resistivity of metastable
  copper-iron solid solutions}, Z.\ Phys.\ B\ Con.\ Mat. \textbf{23}, 23--26
  (1976).
\input{Korn_ZPBCM_1976}

\bibitem{Hihara_JJap_1997}
T.~Hihara, Y.~Xu, T.~J. Konno, K.~Sumiyama, H.~Onodera, K.~Wakoh, and
  K.~Suzuki, \emph{Microstructure and Giant Magnetoresistance in {Fe}-{Cu} Thin
  Films Prepared by Cluster-Beam Deposition}, Jpn.\ J.\ Appl.\ Phys
  \textbf{36}, 3485 (1997).
\input{Hihara_JJap_1997}

\bibitem{Xie_CMS_2011}
Y.-P. Xie and S.-J. Zhao, \emph{The energetic and structural properties of bcc
  {NiCu}, {FeCu} alloys: A first-principles study}, Comp.\ Mat.\ Sci.
  \textbf{50}, 2586--2591 (2011).
\input{Xie_CMS_2011}

\bibitem{James_PRB_1999}
P.~James, O.~Eriksson, B.~Johansson, and I.~A. Abrikosov, \emph{Calculated
  magnetic properties of binary alloys between {Fe}, {Co}, {Ni}, and {Cu}},
  Phys.\ Rev.\ B \textbf{59}, 419--430 (1999).
\input{James_PRB_1999}

\end{thebibliography}

\end{document}